# Targeted and Imaging-guided *In Vivo* Photodynamic Therapy of Tumors Using Dual-functional, Aggregation-induced Emission Nanoparticles

*Xianhe Sun[1,‡], Abudureheman zebibula[2,‡], Xiaobiao Dong[3], Gonghui Li[2,\*], Guanxin Zhang[3,\*], Deqing Zhang[3], Jun Qian[1], Sailing He[1,4\*]*

[1]State Key Laboratory of Modern Optical Instrumentations, Centre for Optical and Electromagnetic Research, Zhejiang University, Hangzhou, Zhejiang, 310058, China

[2]Department of Urology, Sir Run Run Shaw Hospital, College of Medicine, Zhejiang University, Hangzhou, Zhejiang, 310016, China

[3]Beijing National Laboratory for Molecular Sciences, CAS Key Laboratories of Organic Solids and Analytical Chemistry for Living Biosystems, Institute of Chemistry, Chinese Academy of Sciences, Beijing 100190, China

[4]School of Electrical Engineering, Royal Institute of Technology, OSQULDAS VÄG 6, SE-100 44 Stockholm, Sweden

‡These authors contributed equally

**Abstract:** Imaging-guided photodynamic therapy (PDT) has been regarded as a promising strategy for precise cancer treatment. Because of their excellent modifiability and drug loading capacity, nanoparticles have played an important role in PDT. However, when traditional photosensitizers are made into nanoparticles, both their fluorescence and reactive oxygen species (ROS) generation efficiency are decreased due to a phenomenon known as aggregation-caused quenching. Fortunately, in recent years, several kinds of organic dyes have been developed with "abnormal" properties termed aggregation-induced emission (AIE). With enhanced fluorescence emission in the nano-aggregation state, the traditional obstacles mentioned above could be overcome by AIE luminogens (AIEgens). Herein, we provide a better combination of photosensitizers and nanoparticles, a kind of dual-functional AIE nanoparticle capable of producing ROS, to achieve targeted and imaging-guided *in vivo* PDT. Good contrast in *in vivo* imaging and obvious therapeutic efficiency were realized with a low dose of AIE nanoparticles as well as a low power density of light, resulting in negligible side effects. Our work demonstrates that AIE nanoparticles could play a promising role for imaging-guided clinical photodynamic cancer therapy in the near future.

# 1 Introduction

Cancer has increasingly become a primary threat to human health, and more efficient treatment methods are urgently needed[1-5]. Conventional cancer treatment methods, including surgery, chemotherapy, and radiotherapy, lack accuracy and have significant side effects[6-10]. Thus, much attention has been paid to the development of alternative novel treatment modalities[11-15]. Imaging-guided photodynamic therapy (PDT) has been developed and shown to be an efficient, precise, and non-invasive medical technique for cancer therapy[16-20]. The photodynamic effect consists of three elements: a photosensitizer (PS), light of a suitable wavelength, and oxygen. Briefly, the energy of the light can be utilized to transfer the non-toxic triplet oxygen to toxic reactive oxygen species (ROS). Because the process occurs only when both the PS and light of a particular wavelength are present together, this method can provide good selectivity for the treatment region[21-23]. Meanwhile, most PSs demonstrate fluorescence emission[23, 24], and with the guidance of fluorescence images, PDT can be a rather precise treatment modality.

Along with the development of nanotechnology, various series of colloidal nanoparticles have become powerful tools in PDT[25-28]. Since the nanometer scale provides a high surface-to-volume ratio, nanoparticles can ensure high drug-loading capacity as well as efficient surface chemical modification. A high drug-loading capacity leads to fewer side effects, while surface chemical modification allows for customized designs, which can greatly improve the specific properties of drugs, such as hydrophilicity[29], targeting ability[30-33] etc. In addition, the "enhanced permeability and retention" effect, a characteristic of tumor tissue, allows nanoparticles to pass through loose vascular tissue and accumulate in the tumor tissue, providing a passive way of targeting drug delivery[34, 35]. However, for most PSs, especially widely-used porphyrin derivatives, π-π stacking occurs when they are made into nanoparticles, due to their hydrophobic and rigid planar structures. This drawback can result in aggregation-induced quenching of fluorescence[36] and obvious reduction in ROS production[37], which limits the performance of optical imaging and the efficiency of imaging-guided PDT.

Aggregation-induced emission (AIE), an "abnormal" effect discovered in 2001, could provide a solution to the dilemma mentioned above[38-40]. Organic propeller-shaped AIE luminogens (AIEgens), such as tetraphenylethene (TPE) and silole derivatives, are non-emissive or weakly-emissive in benign solvents (usually organic solvents) but become much brighter when forming aggregates in poor solvents (e.g. water)[41]. Fabricating AIEgens into nanoparticles takes advantage of their nature, thus changing the obstacles into opportunity and further benefiting photodynamic therapy of tumors. In the field of photodynamic therapy, AIEgens have been reported to be combined with commonly used PSs, such as PpIX, acting as a sensing probe[42] or an enhancer for ROS production[43, 44]. AIE nanoparticles have also been used only as a PS (not a fluorophore) to achieve *in vivo* PDT via intratumoral injection[45]. In these studies, AIEgens were used as a sensing probe, an energy donor, or just a PS. The procedures were complex and the potential of AIE materials has only been partially utilized. In fact, AIE nanoparticles can be a good photosensitizer as well as a fluorescent label. Recently, some AIEgens with the ability to produce ROS themselves have been synthesized and used for *in vitro* PDT treatment[46, 47]. Very recently, a work reported attempts to achieve imaging-guided and targeted *in vivo* PDT using AIE nanoparticles alone in a dose of 30 mg/kg[48]. While in the present independent work, we have achieved targeted *in vivo* imaging and a similar anti-tumor efficiency (tumor inhibition calculated as 60.4%) with a dose as low as 10 mg/kg, which may induce less biological toxicity and side-effects.

Herein, we fabricated a kind of dual-functional AIE nanoparticle based on a widely-used AIEgen, 2-((4-(2,2-bis(4-methoxyphenyl)-1-phenylvinyl)phenyl)(phenyl)methylene) malononitrile (TPE-red), which can be easily synthesized and bears the outstanding ability to produce ROS[49, 50]. To enhance tumor targeting, promote endocytosis, and ensure a therapeutic effect, we modified the nanoparticles with a customized peptide, cyclic (Arg-Gly-Asp-d-Phe-Cys) (c(RGDfc)), a group commonly used to target over-expressed integrin $\alpha_v\beta_3$ in most cancer cells[51]. As illustrated in **Scheme 1B**, the polyethylene glycol (PEG) and RGD-modified TPE-red nanoparticles (TPE-red-PEG-RGD) were intravenously injected into tumor-bearing mice. It can target the tumor tissues in both passive and active ways, be internalized by cancer cells, and finally produce ROS when subjected to suitable irradiation. Compared with common PSs, the AIE nanoparticles we fabricated avoid the aggregation-caused quenching effect. Taking advantage of bright fluorescence emission, high targeting efficiency, and outstanding ability of ROS production, targeted imaging of tumor regions and high anti-tumor efficiency could be achieved in a rather low dose, which may cause fewer side effect. Furthermore, TPE-red-PEG-RGD is rather simple and easy to make compared with composite structures, which can facilitate further research and mass production.

## 2 Materials and Methods

### 2.1 Materials

All reagents and solvents were purchased from commercially-available sources and used without further purification. 1,2-Distearoyl-*sn-glycero*-3-phosphoethanolamine-N-[maleimide (polyethylene glycol)-2000] (DSPE-PEG$_{2000}$-Mal) was purchased from JenKem Technology Co., Ltd. 2- ((4- (2,2-bis(4-methoxyphenyl)-1-phenylvinyl) phenyl)

(phenyl)methylene) malononitrile (TPE-red) was synthesized according to our previous report[49]. Thiolated cyclic (Arg-Gly-Asp-d-Phe-Cys) peptide (c(RGDfc)) was customized from GL Biochem Ltd (Shanghai). 9,10-anthracenediylbis (methylene) dimalonic acid (ABDA) and 3-(4,5-dimethylthiazol-2-yl)-2,5-diphenyltetrazolium bromide (MTT) were purchased from Sigma-Aldrich (Shanghai, China). Minimum essential medium (MEM), dulbecco minimum essential media (DMEM), RPMI 1640 medium, fetal bovine serum (FBS), and trypsin-EDTA solution were purchased from Gibco. Tetrahydrofuran, dimethyl sulfoxide, hydrochloric acid and PBS (1×) were obtained from the Chemical Reagent Department of Zhejiang University. Deionized (DI) water was used in all experiments, with resistivity of 18.2 MΩ/cm.

## 2.2 Preparation of TPE-red-PEG-RGD nanoparticles

Firstly, TPE-red-PEG nanoparticles were fabricated through a modified nanoprecipitation method[52]. Briefly, 0.335 mL of TPE-red solutions in tetrahydrofuran (1 mg/mL) and 0.5 mL of DSPE-PEG$_{2000}$-Mal solutions in tetrahydrofuran (1.2 mg/mL) were mixed gently in a flask (25 mL). The mixture was then dried under vacuum in a rotary evaporator at room temperature. When the tetrahydrofuran was completely removed, 5 mL of PBS (1×) was added to the flask, and the solution was sonicated for several minutes to form an optically-clear suspension.

To synthesize TPE-red-REG-RGD nanoparticles, the pH value of the as-prepared TPE-red-PEG nanoparticles solution was adjusted to 5.0~7.0, and then 2 mg c(RGDfc) powder was added in a nitrogen atmosphere to start the reaction. The mixture was stirred overnight to make them completely react, and the reaction product was washed with deionized water by dialyzing for two days. After that, the obtained solution was filtered with 0.22 μm microporous membrane, concentrated by vaporizing the water, and finally stored at 4 ℃ for further use.

**2.3 Characterization**

Transmission electron microscopy (TEM) images were taken by a JEOL JEM-1200 transmission electron microscope operating at 80 kV in bright-field mode. The absorption spectra of the nanoparticles were measured with a Shimadzu 2550 UV-vis scanning spectrophotometer. The hydrodynamic size distribution of the nanoparticles was measured on a Malvern Zetasizer Nano ZS-90.

**2.4 Fluorescence spectra measurement**

As illustrated in **Figure S1**, a lab-built fluorescence detection system was used to measure the one-photon and two-photon fluorescence spectra. Briefly, samples in a cuvette were excited by a focused laser beam, and the fluorescence signals were collected laterally by an objective lens (20×, NA = 0.75) and recorded with a spectrometer. In the case of one-photon excited fluorescence, a 450-nm semiconductor laser was used as the excitation source, and the spectra were recorded with a spectrometer of PG 2000, Ideaoptics. In the case of two-photon excited fluorescence, a 1040-nm fs laser [from an amplified output of a large-mode-area ytterbium-doped photonic crystal fiber (PCF) oscillator (1040 nm, 150 fs, 50 MHz)][53]was used as the excitation source, and the spectra were recorded with a spectrometer of QE 6500, Ocean Optics.

**2.5 Power dependence of Two-photon fluorescence**

Series of spectra according to different excitation powers were obtained using the method described above. Based on the spectra, the intensity of fluorescence was calculated as the integral of spectrum envelopes from 500 nm to 670 nm. Then we made a scatter plot of fluorescence intensity and the square of incident power, and fitted the points linearly to verify the linear relationship between them.

**2.6 Reactive oxygen species (ROS) detection *ex vivo***

ABDA was used to evaluate the production of ROS from TPE-red-PEG-RGD. A 200 μL of ABDA solution in DMSO was added into a 2 mL aqueous dispersion of TPE-red-PEG-RGD (50 μg/mL). The mixture was mixed evenly and irradiated with a 450-nm semiconductor laser. The absorption spectra of the mixture were recorded after laser irradiation at various time points, and the decrease of absorption value at 378 nm was used as the indicator of ROS. The same mixture without laser irradiation was used as a control.

**2.7 *In vitro* experiments**

**2.7.1  Cell culture**

UMUC3 cells (human bladder cancer cell line), Hela cells (human cervical cancer cell line), and A549 cells (human pulmonary carcinoma cell line) were obtained from the Cell Culture Center of the Institute of Basic Medical Sciences, Chinese Academy of Medical Sciences (Shanghai, China). The UMUC3 cells, Hela cells, and A549 cells were grown in minimum essential medium (MEM), dulbecco minimum essential media (DMEM), and RPMI 1640 medium, respectively. The culture media were all supplemented with 10% FBS (fetal bovine serum), 1% penicillin, and 1% amphotericin B. The environment was kept at 37 ℃ with 5% $CO_2$.

**2.7.2  *In Vitro* cell imaging**

We used the nanoparticle-treated cells to verify the ability of nanoparticles to be swallowed by the cells. One day before the treatment, UMUC3 cells, Hela cells and A549 cells were seeded in 35-mm cultivation dishes at a confluence of 50 to 60%. During the treatment, the cells were incubated with the appropriate concentration of TPE-red-PEG-RGD nanoparticles for about 2 hours. Cells treated with PBS were used as a control. Afterwards, the cells were all washed thrice with PBS and directly imaged using a microscope.

Two photon excited fluorescence was utilized to achieve *in vitro* cell imaging. As illustrated in **Figure S4**, the imaging system consisted of an upright scanning microscope (Olympus, BX61+FV1200) equipped with a 1040 nm fs laser [from an amplified output of a large-mode-area ytterbium-doped photonic crystal fiber (PCF) oscillator (1040 nm, 150 fs, 50 MHz)][53]. The 1040 nm fs laser beam was guided into the upright scanning microscope and focused onto the cell samples by a 60×/1.00 (Olympus) or a 20×/0.75 (Olympus) water-immersed objective lens to achieve two-photon excitation. The signals with an integration time of 10 μs per pixel were epi-collected with the same objective lens. After passing through a 590-nm long pass filter, a 570-nm dichroic mirror and a 570 nm-625 nm band-pass filter, the fluorescence signals were finally detected by a photomultiplier tube (PMT) via non-descanned detection (NDD) mode.

### 2.7.3 Dark cytotoxicity assay

*In vitro* cytotoxicity was measured by performing MTT assays on UMUC3 cells. Cells were seeded into a 96-well cell culture plate at $5\times10^3$/well and cultured at 37 ℃ with 5% $CO_2$ for 24h. 200 μL of fresh MEM with different concentrations of TPE-red-PEG-RGD (5, 10, 20, 50 μg/mL, diluted in MEM) were then added into the wells. The cells were subsequently incubated for 48 hours at 37 ℃ with 5% $CO_2$. Then, MTT (20 μL/well, 5 mg/mL) was added to each well, and the plate was incubated for an additional 4 hours at 37 ℃ with 5% $CO_2$. The medium was then replaced with 200 μL dimethyl sulfoxide (DMSO) per well, and OD570 was monitored by an enzyme-linked immune sorbent assay (ELISA) reader.

### 2.7.4 Cytotoxicity assay of PDT

Cells were seeded into a 96-well culture plate at $5\times10^3$/well and cultured at 37 ℃ with 5% $CO_2$ for 24h. 200 μL fresh MEM with different concentrations of TPE-red-PEG-RGD (10, 20, 50 μg/mL, diluted in MEM) were then added into the wells. After incubation of 2 hours, the nanoparticles-treated cells were irradiated by 450 nm light (40mW, 200mW/cm$^2$) for 6 min per well. A control group was irradiated by light, but incubated without nanoparticles. Afterwards, the cell viability was measured after incubating for 48 hours with the method introduced above.

## 2.8 *In vivo* experiments

All *in vivo* experiments were performed in compliance with Zhejiang University Animal Study Committee's requirements for the care and use of laboratory animals in research. The animal housing area (located in Animal Experimentation Center of Zhejiang University) was maintained at 24 ℃ with a 12 h light/dark cycle, and animals were fed with water and standard laboratory chow.

### 2.8.1 Tumor xenograft models

UMUC3 tumor model was established by the subcutaneous injection of UMUC3 cells ($5\times10^6$ cells/mL) into the selected positions of the male nude mice (5 weeks old, purchased from Slaccas Co, Ltd (Shanghai), Chinese Academy of Science). To determine tumor size, the greatest length and width of the tumors were determined using a Vernier caliper. The tumor volume was calculated as tumor volume = length × width$^2$ × 0.5.

### 2.8.2 *In vivo* fluorescence imaging

The *in vivo* fluorescence imaging was performed using Iris Spectrum (Perkin Elmer). When tumor volume reached about 50 mm$^3$, 200 μL dispersion of TPE-red-PEG-RGD in 1× PBS (1 mg/mL) was intravenously injected into the mice. The fluorescence imaging was performed at 24, 48, and 72 h post-injection. Spectra of fluorescence signals and autofluorescence signals were picked from the tail in the experimental group and from the back in the control group, respectively. Afterwards, region-of-interests (ROI) were circled around the tumor, and the fluorescence intensities were analyzed by Living Image® Software 4.4. A changing tendency of the amount of TPE-red-PEG-RGD accumulated in the tumor site was plotted as a time-varying line chart.

### 2.8.3  *In vivo* PDT treatment

When tumor volume reached about 50 mm$^3$, 16 tumor-bearing mice were randomly divided into four groups. The treatment scheme was as follows: (1) PBS, without irradiation; (2) PBS, with irradiation; (3) TPE-red-PEG-RGD (10 mg/kg), without irradiation; (4) TPE-red-PEG-RGD (10 mg/kg), with irradiation. The photo-irradiation was applied at 1, 2, and 3 days after the injection (450 nm, 200mW/cm$^2$, for 20 min). The tumor sizes and body weights were inspected every other day for the first 4 days and every day afterwards. The tumor growth inhibition rate was calculated using the following formula:

Tumor growth inhibition rate = (1 – average volume of PDT treated tumors / average volume of tumors in control group) ×100%

The observation lasted for 14 days. Thereafter, the mice were sacrificed, and the tumor masses were weighted and then collected together with major organs for H&E staining.

### 2.9  Histological Examinations

The collected tumors and major organs were fixed in 10% formalin, embedded in paraffin, sectioned, and stained with hematoxylin and eosin. The histological sections were imaged under an inverted optical microscope for analysis.

**Statistics**

All results presented are mean ± s.d. Statistical analysis was performed using Student's t-test. (*$P$< 0.05, **$P$< 0.01, and ***$P$< 0.001).

# 3 Results and Discussion

## 3.1 Synthesis and characterizations of TPEred-PEG-RGD

We synthesized TPE-red according to our previous report[49]. The entire procedure to fabricate our nanoparticles was illustrated as **Scheme 1A**. Firstly, we encapsulated TPE-red with a kind of biocompatible amphiphilic polymer 1,2-Distearoyl-*sn-glycero*-3-phosphoethanolamine-N- [maleimide (polyethylene glycol)-2000] (DSPE-PEG$_{2000}$-Mal)[54, 55], recorded as TPE-red-PEG. The long PEG chains were used to reduce the phagocytosis of nanoparticles by the reticuloendothelial system (RES) [56, 57]. We then modified the target moiety c(RGDfc) through click chemistry between –SH and maleimide at the surface of TPE-red-PEG, yielding TPE-red-PEG-RGD. The morphology and the cross-section structure of TPE-red-PEG-RGD were first checked by transmission electron microscopy (TEM) (**Figure 1A**). The average size of the nanoparticles was about 50 nm, which was confirmed by dynamic light scattering (DLS) (insert in **Figure 1A**). The absorption spectra of TPE-red in tetrahydrofuran (THF) solution and TPE-red-PEG-RGD in PBS are shown in **Figure 1B**. We find that, compared with the absorption spectrum of TPE-red, the spectrum of TPE-red-PEG-

RGD keeps the same shape but has a slight red-shift in wavelength due to the encapsulation process. The fluorescence spectra were recorded using a lab-built fluorescence detection system (**Figure S1**). We adjusted the absorption of both TPE-red and TPE-red-PEG-RGD to the same value and measured the spectra under the same conditions. As can be seen from **Figure 1C**, with the same location of the peak at 650 nm, the fluorescence of TPE-red-PEG-RGD is much stronger than that of TPE-red, which confirms the AIE nature of the materials. Meanwhile, we also compared the fluorescence emitting ability of TPE-red-PEG-RGD and nanoparticles of a common PS, chlorin e6. **Figure S2** shows that, with the same concentration of 100 μg/mL, the fluorescence of TPE-red-PEG-RGD was obviously stronger than that of chlorin e6 nanoparticles, which indicates the advantage of TPE-red-PEG-RGD over traditional PSs. We then verified the ability of TPE-red-PEG-RGD to produce ROS by using a common ROS probe, 9,10-anthracenediylbis (methylene) dimalonic acid (ABDA)[58]. Under irradiation of 450 nm continuous wave (CW) laser light, when mixed with TPE-red-PEG-RGD, the absorption of ABDA at 377 nm kept decreasing over 40 minutes and dropped to about 70 % of the original value (**Figure 1D**, **Figure S3C**). At the same time, almost no change could be observed in control groups (**Figure 1D**, **Figure S3A**, **S3B**). This phenomenon indicated that the TPE-red-PEG-RGD has strong ability to produce ROS continuously under light irradiation.

### 3.2 *In Vitro* cell imaging

The targeting and endocytosis ability of TPE-red-PEG-RGD were verified by two-photon excited fluorescence cell imaging. We first studied the two-photon excited fluorescence properties of TPE-red-PEG-RGD; the spectra were recorded using our lab-built system (see the "Methods" section and **Figure S1**, **Figure 2A**), and power dependence fitting furtherly confirmed the nonlinear optical process (**Figure 2B**). We utilized a scanning microscope to achieve two-photon excited fluorescence imaging. The signal was excited by a 1040 fs laser [from an amplified output of a large-mode-area ytterbium-doped photonic crystal fiber (PCF) oscillator (1040 nm, 150 fs, 50 MHz)][53] and collected by an objective lens. **Figure 2C** and **2D** show the cell images of UMUC3 cells (human bladder cell line) using a 20× objective lens without and with TPE-red-PEG-RGD (0.2μg/mL), respectively. Almost all of the cells were labeled. We also performed cell imaging with a 60× objective lens to verify the label details, and we could clearly find that the signal points distribute well around the nucleus (**Figure 2E**). Furthermore, we compared the endocytosis ability of TPE-red-PEG-RGD using a 20× objective lens with different cell lines (**Figure S5**), which showed the outstanding ability of TPE-red-PEG-RGD on labelling different cancer cells.

### 3.3 MTT assay

A methyl thiazolyltetrazolium (MTT) assay was performed to study the cytotoxicity under dark conditions and the PDT efficiency of TPE-red-PEG-RGD[59]. As shown in **Figure 3A**, TPE-red-PEG-RGD was found to have low toxicity in the dark, and the relative cell viability is kept above 90 % even with a concentration of 50 μg/mL. Upon the irradiation with 450 nm laser light at a power density of 200 mW/cm$^2$, UMUC3 cells without TPE-red-PEG-RGD showed no difference compared to the control group, while the relative cell viability of cells incubated with the nanoparticles dropped below 50 % and decreased as concentration increased, as shown in **Figure 3B**. These phenomena indicate that TPE-red-PEG-RGD is minimally-toxic in the dark and much more toxic under 450 nm light.

### 3.4 *In vivo* fluorescence imaging

To explore the *in vivo* tumor targeting and labeling capability of TPE-red-PEG-ROS, we used tumor-bearing mice as a tumor xenograft model. The UMUC3 cell line was chosen because of its high degree of malignity and representativeness[60, 61]. TPE-red-PEG-RGD in PBS (1×) was intravenously injected, and *in vivo* images were captured at 24h, 48h, and 72h post-injection. We chose a mouse with a strip-shaped tumor to make the label more distinguishable, and the bright field photograph is shown in **Figure 4A**. Fluorescence images at 24h (**Figure 4B**), 48 h (**Figure 4C**), and 72 h (**Figure 4D**) together with the bright field image show that TPE-red-PEG-RGD can label tumor tissues well and last for at least 72 h (**Figure 4F**). The signal from the tumor site was obviously higher than that from peripheral tissue except for the RES (e.g. liver, spleen, etc.) and lymph, and the shape of the labeled area coincidences well with that of the tumor. Spectra taken from the injection port on the tail and the back of the mouse in the control group also confirmed that the fluorescence signal is strongly distinguishable from tissue autofluorescence (**Figure 4E**). With knowledge of organ distributions and metabolic processing of nanoparticles, the liver, spleen, and lymph can be easily recognized, and the additional labeled areas can be determined to be tumor tissue. That will guide us the exact region to apply PDT.

## 3.5 *In vivo* PDT treatment

To further assess the *in vivo* antitumor efficiency of TPE-red-PEG-RGD, 16 tumor-bearing mice were randomly divided into four groups and applied with different treatments: (1) PBS, without irradiation; (2) PBS, with irradiation; (3) TPE-red-PEG-RGD (10 mg/kg), without irradiation; (4) TPE-red-PEG-RGD (10 mg/kg), with irradiation. The observation lasted for 14 days. Thereafter, the mice were sacrificed and photographs were taken of them and the tumors, shown in **Figure 5C** and **5D**, respectively. We recorded the change of tumor volumes. It could be found that only when both TPE-red-PEG-RGD and 450 nm laser light were present could the growth of the tumor volume be obviously inhibited ($P<0.001$) (**Figure 5A**), and the tumor growth inhibition rate was calculated to be as high as 60.4 %. From **Figure 5B**, we could find that the final average tumor weight of the PDT-treated group was significantly lower than the three other control groups ($P<0.05$). The therapeutic efficiency was also confirmed by the tumor tissue sections. Only tumors that were treated with both TPE-red-PEG-RGD and 450-nm-light irradiation have obvious signs of cell death, while nothing of significance happened to tumors in other groups (**Figure 6**).

### 3.6 *In vivo* toxicity

To evaluate the *in vivo* toxicity of TPE-red-PEG-RGD, the weight of the mice was recorded during the treatment. As shown in **Figure S6**, all of mice grew healthily, and not much difference can be observed. We also checked the sections of major organs in different treatment groups; no obvious inflammation or abnormalities could be found, which indicated negligible *in vivo* toxicity of TPE-red-PEG-RGD (**Figure 7**).

## 4 Conclusion

In summary, we demonstrated dual-functional TPE-red-PEG-RGD to achieve targeted and imaging-guided *in vivo* photodynamic therapy. Our work has the following advantages: 1) TPE-red-PEG-RGD takes advantage of the AIE nature of TPE-red molecules and enhance the fluorescence signals instead of suffering from aggregation-caused quenching like many commonly used PSs; 2) both passive and active targeting were utilized to guarantee therapeutic efficiency, and targeted imaging of tumor regions and high anti-tumor efficiency could be achieved with a reasonably low dose, which may cause fewer side effects; 3) the structure of TPE-red-PEG-RGD is simple and easy to fabricate, which could benefit further research and mass production. Our research results show the promising role of AIE nanoparticles in imaging-guided *in vivo* PDT, both in biomedical research and clinical applications.

ACKNOWLEDGMENT

This work was supported by National Basic Research Program of China (973 Program; 2013CB834704 and 2011CB503700), the National Natural Science Foundation of China (11621101), and the Science and Technology Department of Zhejiang Province (2010R50007)

ASSOCIATED CONTENT

**Supplementary materials**. Supplementary figures: Schematic illustration of the lab-built fluorescence detect system; Photograph of aqueous dispersion of TPE-red-PEG-RGD and Chlorin e6 nanoparticles under daylight and under UV lamp; Absorption spectra of ABDA containing solutions with different treatments; Schematic illustration of the two-photon excited fluorescence imaging system; Average body weight variation of the mice during the treatment.

This material is available free of charge *via* the Internet at http://

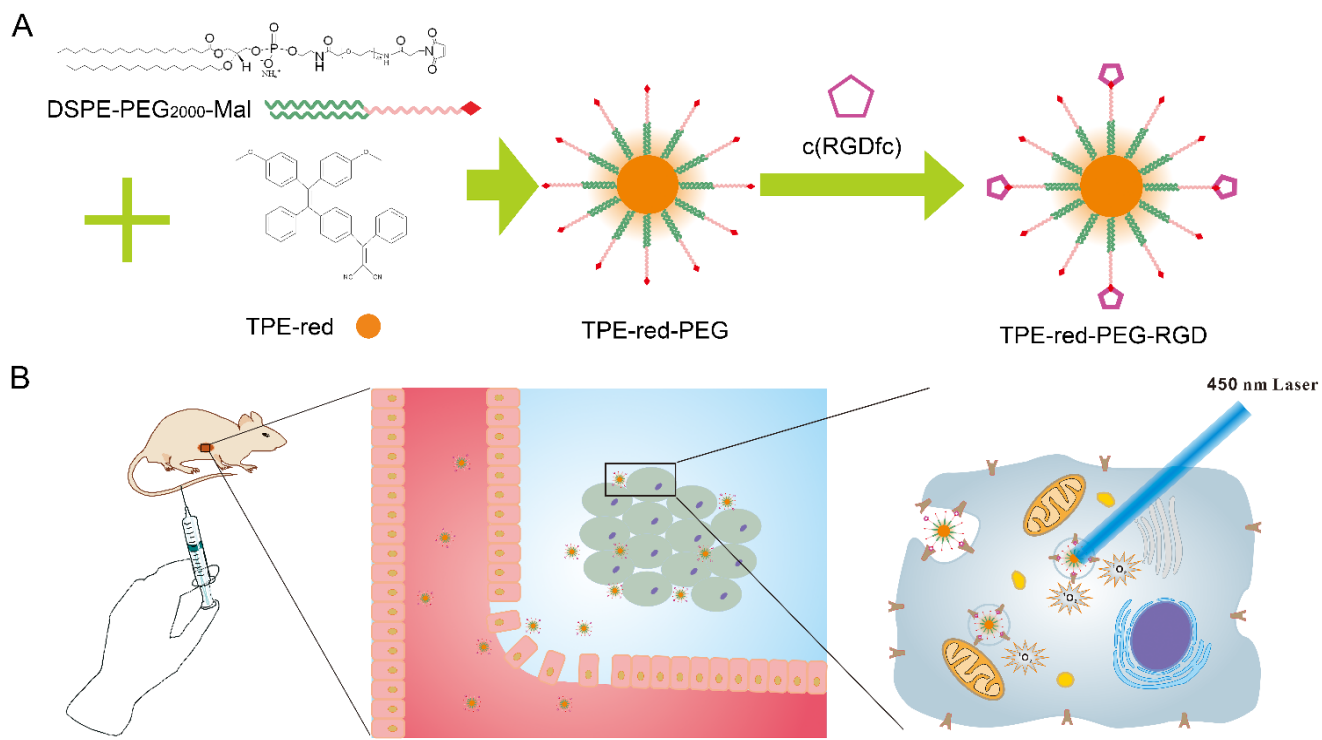

**Scheme 1.** Schematic illustration of (A) the fabrication procedure of TPE-red-PEG-RGD and (B) the process of targeting and image-guided *in vivo* photodynamic cancer therapy with intravenous injection.

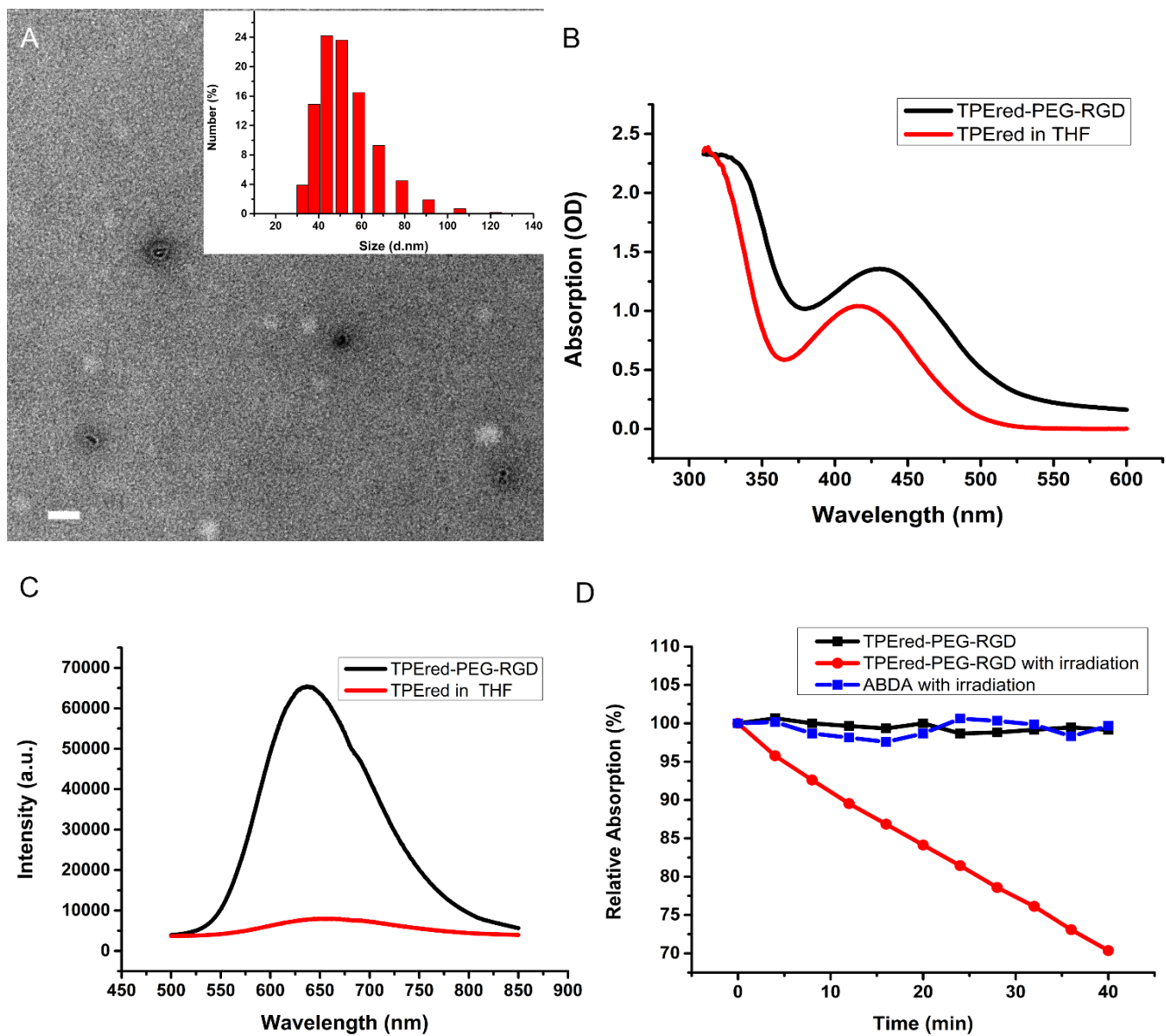

**Figure 1.** Characteristics of TPE-red-PEG-RGD. (A) TEM image of TPE-red-PEG-RGD (insert: DLS data); (B) absorption spectra and (C) fluorescent spectra of TPE-red-PEG-RGD in PBS (1×) and TPE-red in THF; (D) Normalized absorption of ABDA at 377 nm with different treatments.

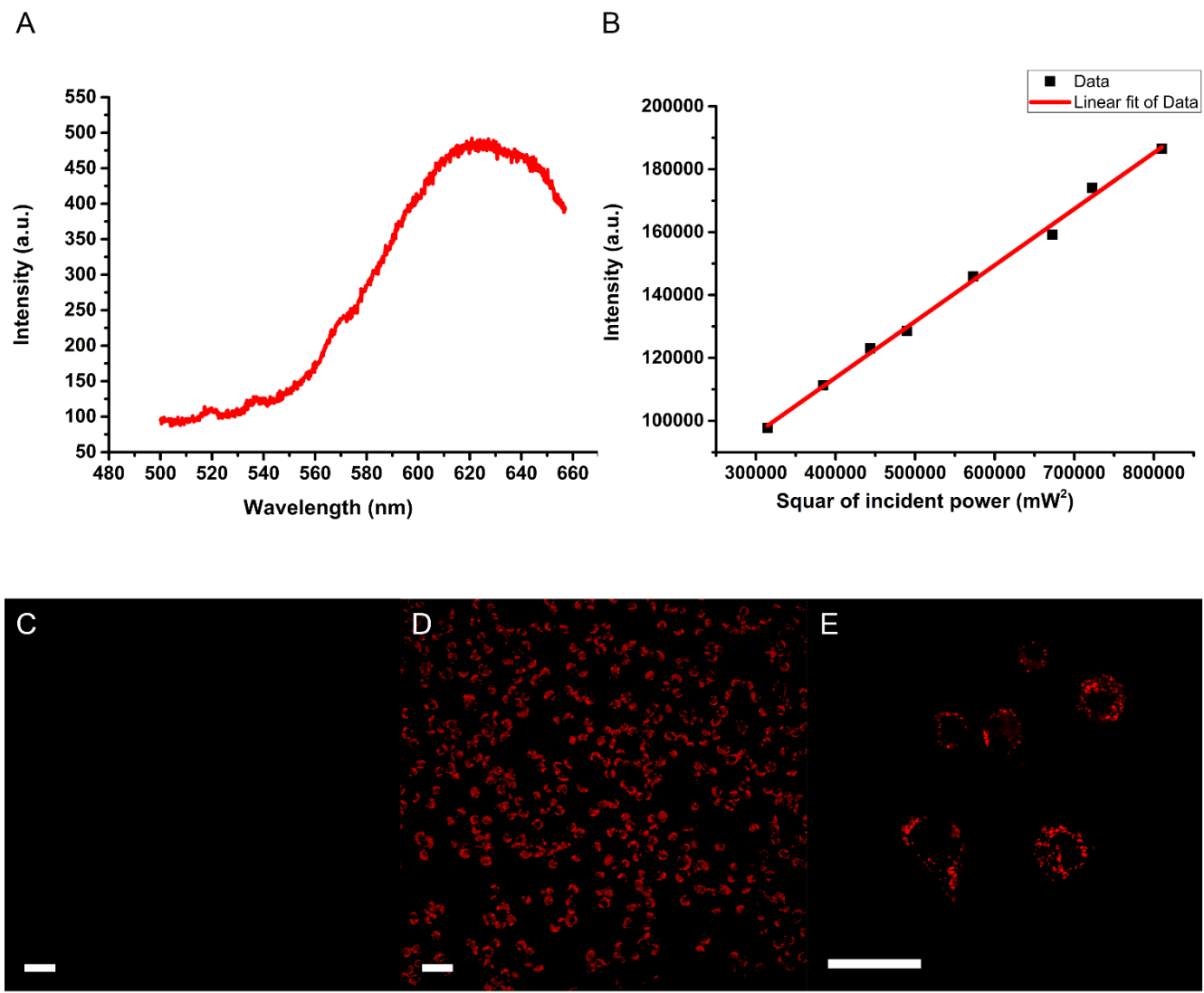

**Figure 2.** Two-photon excitation property and *in vitro* cell imaging of UMUC3 cells incubated with TPE-red-PEG-RGD. (A) Two-photon excited fluorescence spectrum of TPE-red-PEG-RGD; (B) the power dependence relationship and linear fitting of the data; two-photon excited fluorescence images of UMUC3 cells (C) without nanoparticle treatment using a 20× objective lens and (D), (E) treated with TPE-red-PEG-RGD (0.2 μg/mL) using a 20× objective lens and a 60× objective lens, respectively. Scale bar, 50 μm.

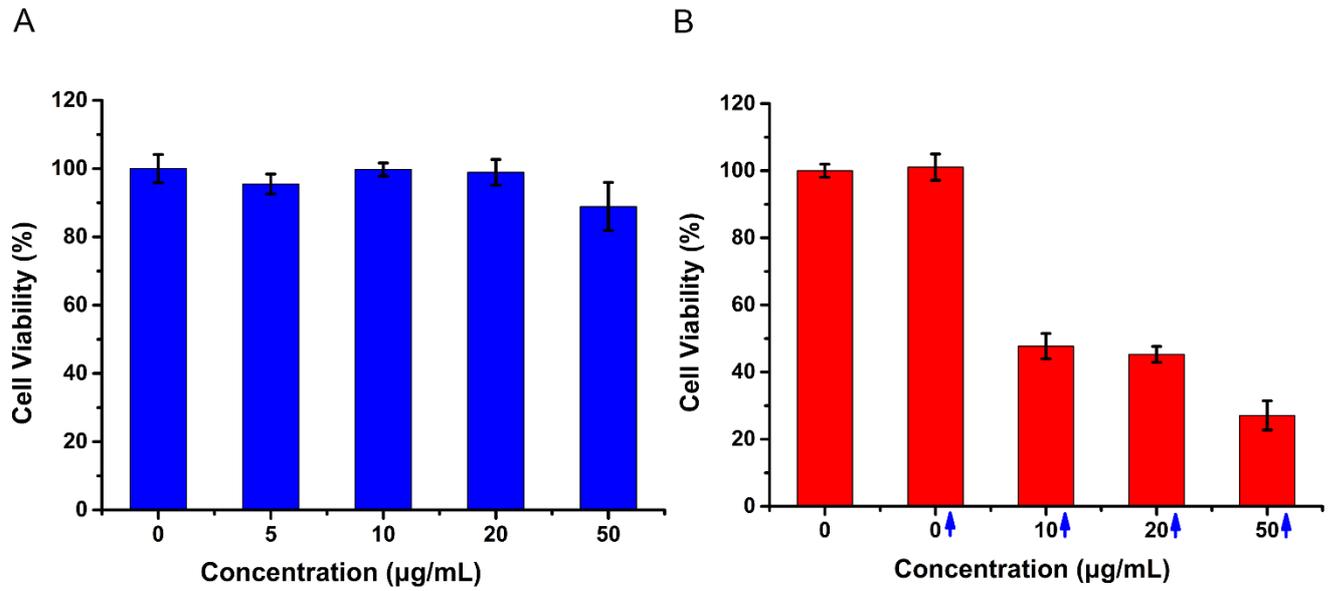

**Figure 3.** Viability of UMUC3 cells treated with various concentrations of TPE-red-PEG-RGD, (A) without and (B) with irradiation. The blue arrows indicate the presence of light. Error bars indicate SD.

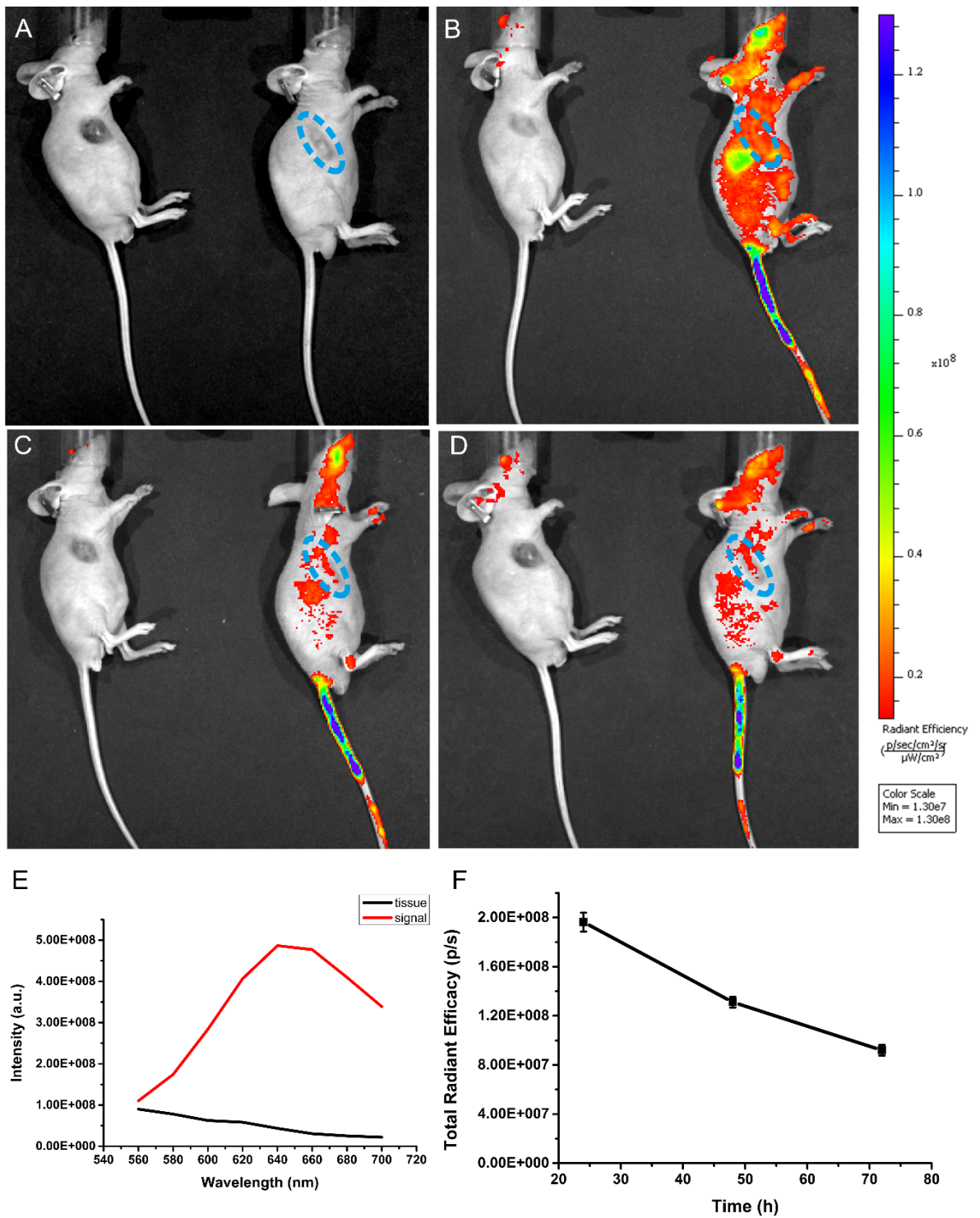

**Figure 4.** *In vivo* imaging studies of tumor bearing mice treated with TPE-red-PEG-RGD. (A) Photograph of mice in the control group (left) and experimental group (right); (B)-(D) fluorescence images of mice 24, 48, and 72 h post-injection in control group (left) and experimental group (right); (E) spectra of TPE-red-PEG-RGD (taken from injection port on tail) and tissue autofluorescence (taken from the back of a mouse in the control group); (F) Changing tendency of total radiant efficiency in the tumor area (circled in blue). Error bars indicate SD.

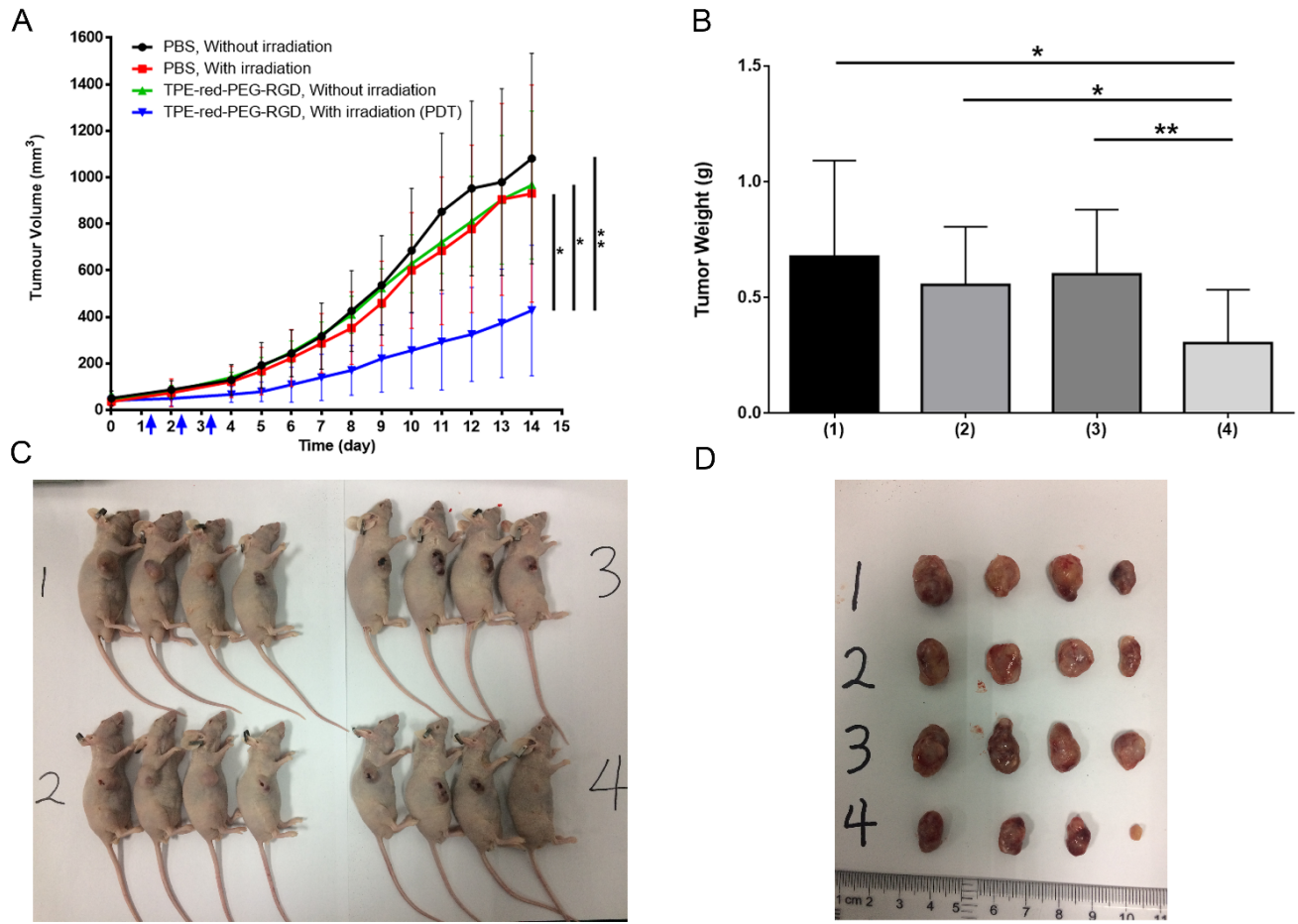

**Figure 5.** *In vivo* anti-tumor study of TPE-red-PEG-RGD. (A) Change of average tumor volume with different treatments as time passed (blue arrows indicate the presence of light); (B) Average tumor weights with different treatments at 14 days; Photographs of (C) sacrificed mice and (D) tumors. The numbers written on the paper indicate different groups: 1) PBS, no irradiation; 2) PBS, irradiation; 3) TPE-red-PEG-RGD, no irradiation; 4) TPE-red-PEG-RGD, irradiation (PDT). Error bars indicate SD. Scale bar, 50 μm. (*$P<0.05$, **$P<0.01$)

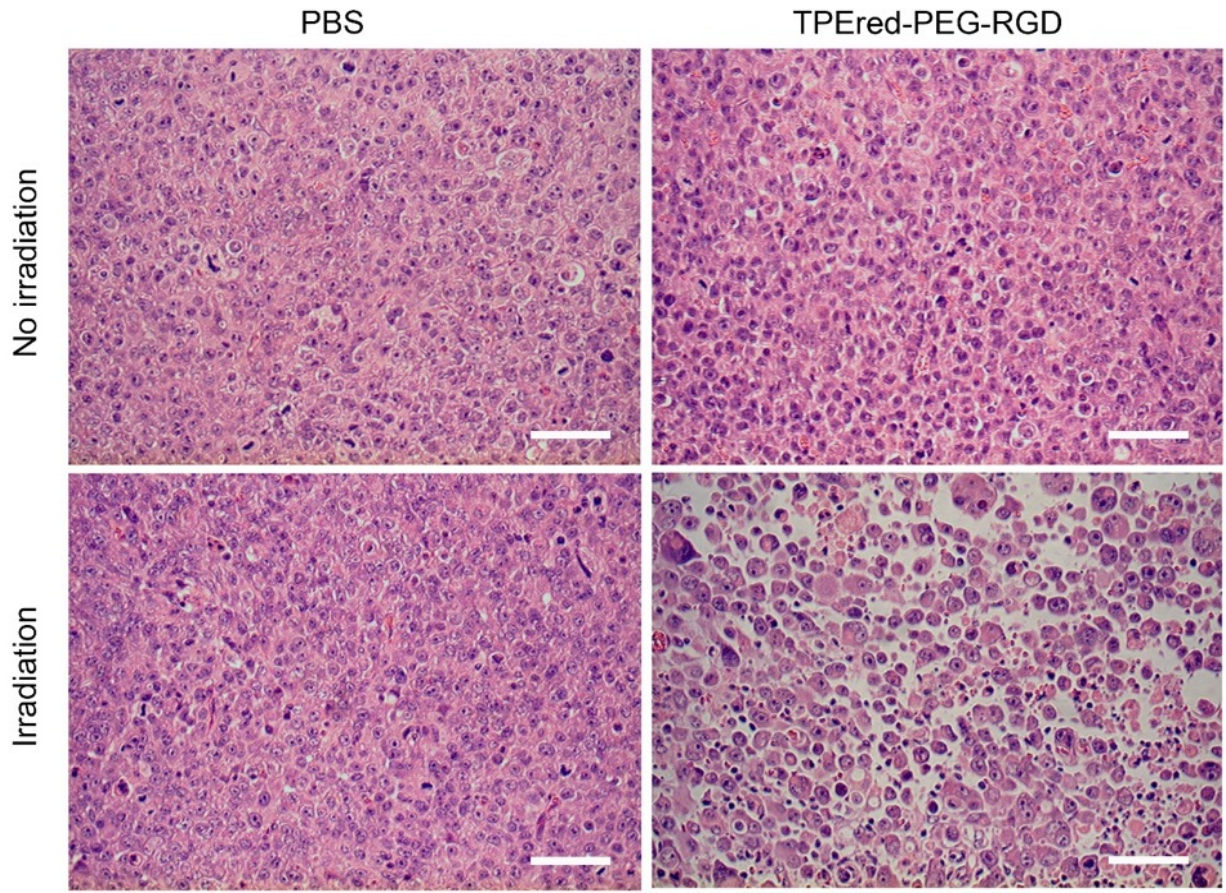

**Figure 6.** H&E stained sections of tumor tissues in different treatment groups. Scale bar, 50 μm.

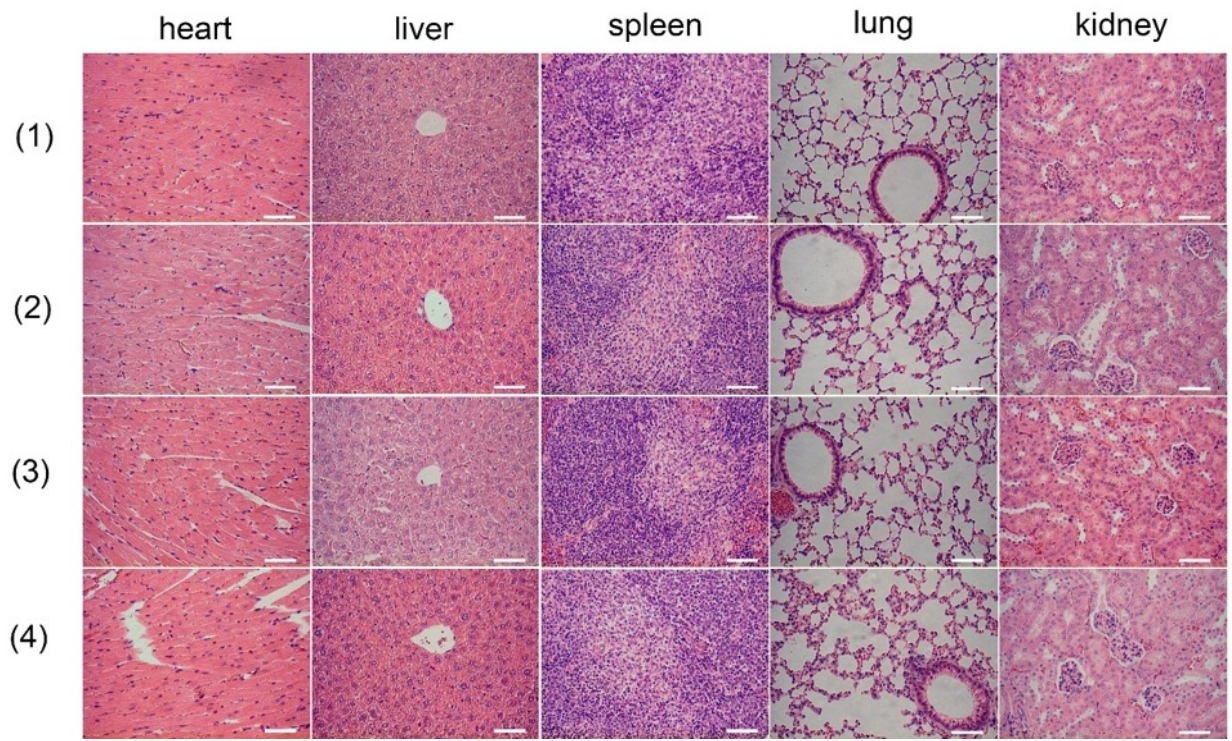

**Figure 7.** H&E stained sections of major organs (heart, liver, spleen, lung, and kidney) of mice with different treatments. The numbers written on the paper indicate different groups: 1) PBS, no irradiation; 2) PBS, irradiation; 3) TPE-red-PEG-RGD, no irradiation; 4) TPE-red-PEG-RGD, irradiation (PDT). Scale bar, 50 μm.